\documentclass[preprint, 12pt]{elsarticle}

\usepackage{url}
\usepackage{amssymb}
\usepackage{amsmath}
\usepackage{booktabs}
\usepackage{graphicx}
\usepackage{subcaption}
\usepackage{multirow}
\usepackage{fixltx2e}
\usepackage{rotating} 
\usepackage{wasysym}

\journal{Medical Image Analysis}

\begin{document}

\begin{frontmatter}



\title{Benchmarking and Improvement of Artifact Segmentation Methods in H\&E-Stained Whole Slide Imaging} 


\author[1]{Julia Merta\corref{cor1}}
\ead{julia.merta@polsl.pl}

\author[2]{Daniel Mendelsohn}

\author[2]{Benedikt Schoser}

\author[1]{Joanna Polanska}

\author[1,3]{Michal Marczyk\corref{cor1}}
\ead{michal.marczyk@polsl.pl}
\cortext[cor1]{Corresponding authors}

\affiliation[1]{
    organization={Department of Data Science and Engineering,
                  Silesian University of Technology},
    addressline={Akademicka 16},
    city={Gliwice},
    postcode={44-100},
    country={Poland}
}

\affiliation[2]{
    organization={Friedrich-Baur-Institute, Department of Neurology, LMU Medizin},
    city={Munich},
    postcode={80336},
    country={Germany}
}

\affiliation[3]{
    organization={Yale Cancer Center, Yale University},
    city={New Haven},
    postcode={06511},
    state={CT},
    country={USA}
}


\begin{abstract}
Technical artifacts introduced during tissue preparation or slide scanning remain a major obstacle to developing robust computational pathology algorithms for hematoxylin and eosin (HE)-stained whole-slide images (WSIs). Despite the growing number of quality-control tools, their real-world utility remains hard to assess because of the limited number of publicly available implementations and the lack of systematic, multi-dataset comparisons.
This paper presents a benchmark of artifact segmentation methods for HE-stained WSIs. After a literature review, six publicly available algorithms were selected and compared in terms of their ability to segment artifact-free tissue and selected artifact classes. Two publicly available datasets and a newly introduced dataset released to support further research were used for evaluation. Tissue samples from multiple medical centers and different organs are included, enabling evaluation across heterogeneous data sources. Segmentation performance was assessed using Dice Index, pixel-level recall, and false discovery rate for artifact-free tissue segmentation. For algorithms supporting multiclass artifact segmentation, class-wise Dice Index and inter-class misclassification were evaluated. In addition to segmentation quality, processing time was assessed.
Statistical analysis revealed significant differences in segmentation performance between methods. HistoKit, which incorporates a modified version of the GrandQC, was identified as the best-performing algorithm, while HistoQC was the worst. In terms of artifact types, dark spots and out of focus regions were harder to segment than others. Finally, a simple post-processing step was introduced for HistoKit, which further improved segmentation performance.
The benchmark establishes a reference for evaluating and developing artifact segmentation methods for HE-stained WSIs.

\end{abstract}


\end{frontmatter}



\section{Introduction}

Whole Slide Imaging (WSI), introduced in 1999, enables the creation of high-resolution digital slides of tissue specimens, supporting the development of extensive public repositories for educational and research purposes \citep{zuraw_whole-slide_2022}. This advancement greatly facilitates slide storage and transfer between institutions, which speeds up collaboration among pathologists at medical centers worldwide \citep{verghese_computational_2023}. As computing power increased and newer scanners entered the market, computational pathology (CPath) has emerged as a discipline focused on analyzing histological data through machine learning and image processing algorithms \citep{jahanifar_domain_2025}. Recently, numerous deep learning-based methods have been developed for the automated analysis of WSI to assist physicians and speed up the diagnostic process \citep{cui_artificial_2021}. Models based on convolutional neural networks (CNNs) and vision transformers (ViTs) represent state-of-the-art approaches for tasks such as cell or tissue segmentation, prediction of various clinical characteristics of a tissue, or survival risk regression \citep{10190115, Cooper2023}.

Currently, major challenges stand in the way of the widespread adoption of CPath in clinical practice. Although existing methods achieve excellent results on distinct datasets, they often perform poorly on unseen data, posing a significant barrier to their implementation in real-world environments \citep{verghese_computational_2023, jahanifar_domain_2025}. Moreover, the effectiveness of deep learning models is highly dependent on the representativeness of the training dataset \citep{tellez_quantifying_2019, hosseini_computational_2024}. Therefore, successful deployment requires models with good generalization capabilities that can accommodate the variability present in histology data. This heterogeneity arises from multiple sources, but mostly from differences in staining protocols and different types of artifacts \citep{jahanifar_domain_2025}.

Artifacts are unwanted structures or inconsistencies that obstruct clear visualization of tissue morphology \citep{taqi_review_2018}. They may be introduced during any stage of WSI preparation. \citet{Kanwal2022} classified WSI artifacts into three primary categories: tissue-level, slide-level, and scan-level artifacts.

\begin{figure*}[t!]
    \centering
    \includegraphics[width=1\linewidth]{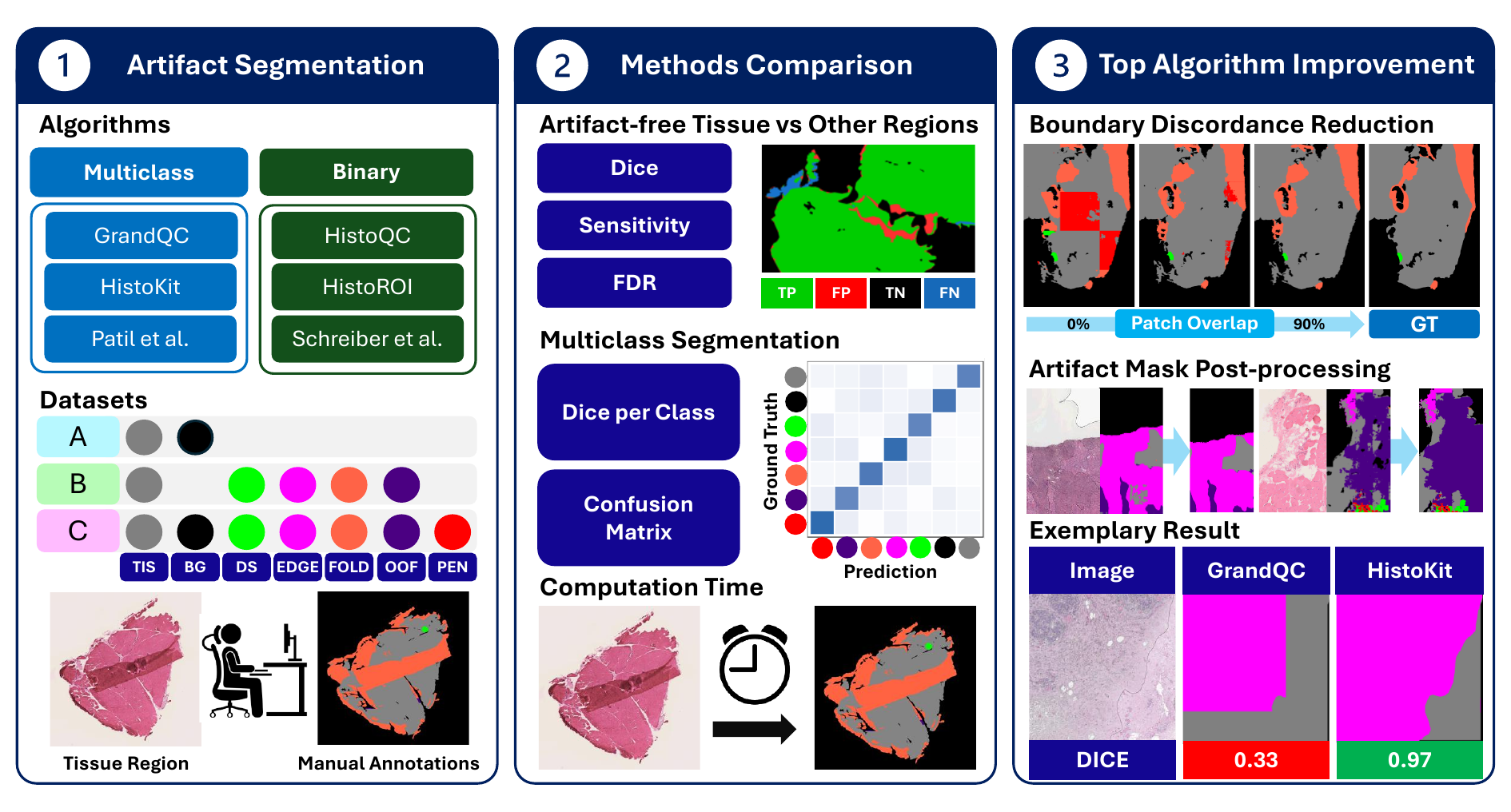}
    \caption{Overview of the study design and algorithm comparison workflow. Six publicly available algorithms: GrandQC \citep{wengGrandQCComprehensiveSolution2024a}, HistoKit \citep{bioimaging26}, HistoQC \citep{janowczyk_histoqc_2019}, HistoROI \citep{patil_efficient_2023}, \citet{patil2025semanticsegmentationbasedquality}, and \citet{schreiber_rapid_2024},  were evaluated based on segmentation quality, artifact classification, and processing time. Three datasets comprising multiple artifact classes were used in this study. The annotated classes included dark spots and foreign objects (DS), air bubbles and other edge-like artifacts (EDGE), tissue folds (FOLD), out-of-focus regions (OOF), and pen-marker traces of different colors (PEN), as well as artifact-free tissue (TIS) and background (BG). For Dataset A, background and artifact regions were considered jointly. Additionally, the parameters of the boundary discordance reduction for the best-performing method were examined. Moreover, a post-processing method for the artifact mask for the top-performing algorithm to further enhance segmentation quality.}
    \label{fig:pipeline}
\end{figure*}

Tissue-level artifacts arise during tissue preparation, a stage that requires significant precision and expertise in both laboratory procedures and sample collection \citep{lyon_standardization_1994}. The final appearance of the tissue specimen is influenced by numerous variables. Minor variations in preparation protocols, such as fixation time, temperature, tissue thickness, or reagent volumes, are almost inevitable in routine laboratory practice. Therefore, establishing a universal protocol for all laboratories is nearly impossible \citep{Woods1994LaboratoryH}. Several types of tissue-level artifacts can occur during slide preparation. For example, excessive dehydration of the tissue can cause it to shrink. Cracks in tissue may also be seen, most often resulting from a dull knife or an improperly performed cut. Another common finding is tissue folding. Unfortunately, artifacts from this group are often challenging or even impossible to repair. In most cases, this would require repeating the sample collection process or even performing a new biopsy \citep{Kanwal2022}.

Slide-level artifacts can occur in the final stages of the pathology workflow, including slide mounting and storage. These include foreign objects such as dust, hair, and microorganisms, as well as air bubbles introduced during coverslip placement, pen markings in different colors, and dirt. Although marker traces can be removed prior to scanning \citep{Kanwal2022}, they remain common, even in public databases such as The Cancer Genome Atlas (TCGA) \citep{Weinstein2013TCGA}.

Scan-level artifacts occur during tissue scanning, with blur being the most prevalent. This artifact is considered one of the most critical in WSI \citep{shrestha_quantitative_2016}. Out-of-focus regions may result from improper focal calibration or uneven tissue thickness. Furthermore, certain scanners can introduce distortions along the borders of stitched patches, resulting in strip- or grid-like patterns when illumination is inadequate \citep{9300148}. These artifacts are generally independent of the specimen on the slide. Therefore, removing them simply requires re-scanning the slide \citep{Kanwal2022}. However, this process entails additional labor and time that could otherwise be devoted to other tasks. 

In CPath, artifacts are more common and often have a greater influence on model outcomes than in natural images. This is because differences between tissues or biological conditions are more subtle and harder to detect, meaning that even minor distortions can have a significant impact on the model’s predictions, which may lead to an incorrect diagnosis \citep{zhang_benchmarking_2022}. Consequently, quality inspection is an essential component of the data processing workflow \citep{verghese_computational_2023}. This process enables the analysis to focus on artifact-free regions or to entirely exclude low-quality WSI images from further evaluation. Moreover, this step enables identification of issues in specimen preparation and scanning protocol. This may help determine whether a laboratory can make changes to improve slide quality \citep{wengGrandQCComprehensiveSolution2024a}.

Numerous methods have recently been developed for searching artifacts in WSIs. Nevertheless, there is a lack of systematic comparisons evaluating the effectiveness of these methods across multiple manually annotated datasets using segmentation quality metrics. Furthermore, few publicly available datasets for multiclass artifact segmentation exist. This study addresses these research gaps. Figure~\ref{fig:pipeline} presents an overview of the study workflow.

The main contributions of this work are as follows:
\begin{itemize}
\item We prepared a new dataset for technical artifact segmentation, containing manually annotated labels for multiple artifact classes. The dataset enables the evaluation of segmentation methods at the WSI level and includes images from multiple organs and databases.
\item We performed a literature review and comprehensive comparison of existing methods for artifact segmentation in HE-stained images, along with a statistical analysis of their performance and computation time. To the best of our knowledge, this is the first comparison of this scale to have been conducted.
\item We introduced a simple post-processing method to a leading artifact segmentation algorithm that improves detection of technical artifacts.
\end{itemize}

In our study, we limited the analyzed datasets to those containing HE-stained images, as this staining method is the most widely used in pathology. Furthermore, most currently available artifact detection tools have been developed exclusively for this staining technique.

\section{Artifact Segmentation Methods}

Several methods have been developed for artifact segmentation. Some approaches utilize image metrics such as brightness, contrast, and sharpness, while others employ machine learning techniques, including deep learning. Algorithms can be categorized into those that distinguish artifact-free tissue from background and artifacts, and those that further identify specific artifact types at the pixel or patch level. Additionally, certain tools provide reports or recommendations for subsequent actions, such as re-scanning the slide. However, many of these tools are not publicly available, and often do not provide access to trained model weights or the training datasets, so cannot be used in practice. A comparison of recently published algorithms is provided in Table \ref{tab:algorithms_comparison}. The supplementary materials provide detailed information about the errors that occurred during the launch of the tools.

\begin{sidewaystable*}
\centering
\caption{Summary of available artifact segmentation methods for WSIs.}
\label{tab:algorithms_comparison}
\resizebox{0.9\textheight}{!}{%
\begin{tabular}{lllll}
\hline
\multicolumn{1}{c}{Method} &
  \multicolumn{1}{c}{Staining Types} &
  \multicolumn{1}{c}{Approach} &
  \multicolumn{1}{c}{Output} &
  \multicolumn{1}{c}{\begin{tabular}[c]{@{}c@{}}Code/Model\\ Availability\end{tabular}} \\ \hline
 &
   &
   &
   &
   \\
\multicolumn{1}{c}{HistoQC \citep{janowczyk_histoqc_2019}} &
  \multicolumn{1}{c}{H\&E, PAS, SIL, TRI} &
  \multicolumn{1}{c}{\begin{tabular}[c]{@{}c@{}}Image quality metrics, \\ supervised classifiers, edge detectors\end{tabular}} &
  \multicolumn{1}{c}{\begin{tabular}[c]{@{}c@{}}Tissue segmentation masks \\ after each step and a final report\end{tabular}} &
  \multicolumn{1}{c}{Available} \\
 &
   &
   &
   &
   \\
\multicolumn{1}{c}{PathProfiler \citep{Haghighat_PathProfiler}} &
  \multicolumn{1}{c}{H\&E} &
  \multicolumn{1}{c}{ResNet18 with six outputs} &
  \multicolumn{1}{c}{\begin{tabular}[c]{@{}c@{}}Patch-level quality segmentation masks\\ and overall usability scores\end{tabular}} &
  \multicolumn{1}{c}{Available} \\
 &
   &
   &
   &
   \\
\multicolumn{1}{c}{HistoROI \citep{patil_efficient_2023}} &
  \multicolumn{1}{c}{H\&E} &
  \multicolumn{1}{c}{\begin{tabular}[c]{@{}c@{}}ResNet18 trained with\\ human-in-the-loop training\end{tabular}} &
  \multicolumn{1}{c}{\begin{tabular}[c]{@{}c@{}}Patch-level segmentation, CSV file with patch\\ classification labels, and GeoJSON files\end{tabular}} &
  \multicolumn{1}{c}{Available} \\
 &
   &
   &
   &
   \\
\multicolumn{1}{c}{\citet{smit2021quality}} &
  \multicolumn{1}{c}{\begin{tabular}[c]{@{}c@{}}Eight staining types\\ (including H\&E)\end{tabular}} &
  \multicolumn{1}{c}{\begin{tabular}[c]{@{}c@{}}Resolution-agnostic CNN (tissue), \\ DeepLabV3+ with an \\ EfficientNet-B2 encoder (artifacts)\end{tabular}} &
  \multicolumn{1}{c}{\begin{tabular}[c]{@{}c@{}}Multiclass segmentation mask\\ and recommended action\end{tabular}} &
  \multicolumn{1}{c}{\begin{tabular}[c]{@{}c@{}}Unavailable:\\ missing model weights\end{tabular}} \\
 &
   &
   &
   &
   \\
\multicolumn{1}{c}{SliDL \citep{berman_slidl_2023}} &
  \multicolumn{1}{c}{H\&E} &
  \multicolumn{1}{c}{DenseNet} &
  \multicolumn{1}{c}{\begin{tabular}[c]{@{}c@{}}Three-class segmentation:\\ tissue, artifact, and other\end{tabular}} &
  \multicolumn{1}{c}{\begin{tabular}[c]{@{}c@{}}Unavailable:\\ missing model weights\end{tabular}} \\
 &
   &
   &
   &
   \\
\multicolumn{1}{c}{\citet{schreiber_rapid_2024}} &
  \multicolumn{1}{c}{H\&E} &
  \multicolumn{1}{c}{\begin{tabular}[c]{@{}c@{}}Otsu thresholding in a\\ transformed RGB color space\end{tabular}} &
  \multicolumn{1}{c}{Binary segmentation mask} &
  \multicolumn{1}{c}{Available} \\
 &
   &
   &
   &
   \\
\multicolumn{1}{c}{GrandQC \citep{wengGrandQCComprehensiveSolution2024a}} &
  \multicolumn{1}{c}{H\&E} &
  \multicolumn{1}{c}{\begin{tabular}[c]{@{}c@{}}Tissue and artifact segmentation modules\\ with EfficientNet-B0 encoders\\ and U-Net++ decoders\end{tabular}} &
  \multicolumn{1}{c}{Multiclass segmentation mask} &
  \multicolumn{1}{c}{Available} \\
 &
   &
   &
   &
   \\
\multicolumn{1}{c}{AIRAQc \cite{gautam2025airaqc}} &
  \multicolumn{1}{c}{H\&E, IHC} &
  \multicolumn{1}{c}{\begin{tabular}[c]{@{}c@{}}Multi-magnification transformer for\\ tissue and artifact detection\end{tabular}} &
  \multicolumn{1}{c}{Multiclass segmentation mask} &
  \multicolumn{1}{c}{Unavailable} \\
 &
   &
   &
   &
   \\
\multicolumn{1}{c}{\citet{patil2025semanticsegmentationbasedquality}} &
  \multicolumn{1}{c}{H\&E} &
  \multicolumn{1}{c}{\begin{tabular}[c]{@{}c@{}}Four modules based on U-Net/U-Net++\\ with ResNet34/ResNet18 backbones\end{tabular}} &
  \multicolumn{1}{c}{Multiclass segmentation mask} &
  \multicolumn{1}{c}{Available} \\
 &
   &
   &
   &
   \\
\multicolumn{1}{c}{DiffusionQC \citep{wang2026diffusionqcartifactdetectionhistopathology}} &
  \multicolumn{1}{c}{H\&E} &
  \multicolumn{1}{c}{Fine-tuned PixCell-1024 model} &
  \multicolumn{1}{c}{Multiclass segmentation mask} &
  \multicolumn{1}{c}{Unavailable} \\
 &
   &
   &
   &
   \\
\multicolumn{1}{c}{SlideInspect \citep{scotto_slideinspect_2026}} &
  \multicolumn{1}{c}{H\&E, IHC, PAS, TRIC} &
  \multicolumn{1}{c}{\begin{tabular}[c]{@{}c@{}}K-Net with Swin Transformer backbone\\ and UPerNet encoder-decoder architecture\end{tabular}} &
  \multicolumn{1}{c}{\begin{tabular}[c]{@{}c@{}}Multiclass segmentation mask, staining quality\\ information, and recommended action\end{tabular}} &
  \multicolumn{1}{c}{Unavailable} \\
 &
   &
   &
   &
   \\
\multicolumn{1}{c}{HistoKit \citep{bioimaging26}} &
  \multicolumn{1}{c}{H\&E} &
  \multicolumn{1}{c}{\begin{tabular}[c]{@{}c@{}}Thresholding-based tissue segmentation\\ and improved inference with the GrandQC\\ artifact segmentation module\end{tabular}} &
  \multicolumn{1}{c}{\begin{tabular}[c]{@{}c@{}}Multiclass segmentation mask, class-specific\\ confidence masks, and final report\end{tabular}} &
  \multicolumn{1}{c}{Available} \\
 &
   &
   &
   &
   \\
\multicolumn{1}{c}{GuardQC \citep{wei_tissue-preserving_2026}} &
  \multicolumn{1}{c}{H\&E} &
  \multicolumn{1}{c}{\begin{tabular}[c]{@{}c@{}}GrandQC artifact segmentation module,\\ dual-stream ResNet18, and\\ adaptive Otsu thresholding\end{tabular}} &
  \multicolumn{1}{c}{Binary segmentation mask} &
  \multicolumn{1}{c}{Available on request} \\
 &
   &
   &
   &
   \\
\multicolumn{1}{c}{GrandQC+ \citep{borrecho_grandqc_2026}} &
  \multicolumn{1}{c}{H\&E} &
  \multicolumn{1}{c}{Same as GrandQC} &
  \multicolumn{1}{c}{\begin{tabular}[c]{@{}c@{}}Multiclass segmentation mask\\ and final report in an Excel file\end{tabular}} &
  \multicolumn{1}{c}{Unavailable} \\
 &
   &
   &
   &
   \\ \hline
\end{tabular}}
\end{sidewaystable*}

HistoQC \citep{janowczyk_histoqc_2019} is a widely used open-source tool for WSI quality control. It not only enables the identification of artifact-free regions but also the discovery of outliers in the cohort, e.g., slides with incorrect exposure. It works by combining image metrics such as brightness and contrast, color histograms, and features from edge detectors and supervised classifiers to identify artifacts. HistoQC generates visualizations after each processing step and allows users to customize workflows as needed. Additionally, a comprehensive report for the whole dataset is generated at the end of the analysis. The tool was evaluated on 450 HE-stained WSIs from the TCGA BRCA dataset by two pathologists. The results obtained were in concordance with pathologists' observations in over 95\% of cases. Moreover, HistoQC was used for quality control of WSI kidney biopsies stained with other stains, such as Periodic Acid Schiff (PAS), Silver (SIL), and Trichrome (TRI) staining \citep{Chen}. The source code of HistoQC, along with its documentation, is publicly available at \url{https://github.com/choosehappy/HistoQC}. In this study, HistoQC v2.1 was used in a Docker container with default parameters. It is important to note that the tool requires tissue images to be stored in a pyramidal format, with zoom-level information available in the file metadata. Therefore, it could not be applied to Dataset B, for which the magnification was known but could not be provided as an input parameter. It was assumed that the final mask file (with the\textit{ mask\_use} suffix) represents the tissue without artifacts, and that all other pixels belong to the background or artifacts.

\citet{Haghighat_PathProfiler} proposed a tool developed for automated quality assessment of WSIs called Pathprofiler. This algorithm performs patch-level classification using the ResNet18~\citep{7780459} architecture, with a final fully connected layer producing six class scores through linear output units. The scores correspond to the following classes: usability, no artifact, staining artifacts, focus artifacts, tissue folds, and other artifacts. It also uses patch-level statistics to produce overall usability, out-of-focus, and staining scores for each WSI. Pathprofiler was trained on 198 FFPE HE-stained WSIs, comprising 198 prostate biopsy and transurethral resection specimens from two private cohorts. The tool was further evaluated on unannotated WSIs from 499 TCGA cases containing prostate tissue, including frozen tissue sections, as well as on the FOCUS dataset, which includes 788 WSIs from colorectal cancer specimens. For the FOCUS dataset, quality assessment was restricted to the tumor area. Principal component analysis (PCA) showed that FOCUS WSIs labeled as unusable overlapped with selected examples from TCGA and the annotated test dataset in the principal component space, suggesting that PathProfiler may generalize across different datasets. However, no formal analysis using patch classification or segmentation metrics was performed to substantiate this observation. The tool's code is publicly available at \url{https://github.com/MaryamHaghighat/PathProfiler} (last commit: Jan 4, 2024). PathProfiler also does not support specifying the zoom level as an input parameter for tissue regions saved in the non-pyramidal format. In addition, the program did not execute successfully, likely due to issues with the multithreaded slide-processing implementation. As a result, this tool was excluded from the analysis.

HistoROI \citep{patil_efficient_2023} is a deep learning-based method for WSI quality control and patch-level artifact classification. It uses the ResNet18 architecture, trained with a human-in-the-loop active learning method, to predict one of six classes: epithelium, stroma, lymphocyte, adipose tissue, artifact, and miscellaneous area. The artifact class includes patches with out-of-focus regions, coverslips, tissue folds, air bubbles, staining problems, and marker traces. The model was trained on 50 HE-stained WSIs containing breast cancer tissue. The innovation of this method lies in training the model under human supervision to identify WSI images containing regions that are difficult to classify, and in manually correcting the model’s predictions during training to improve the accuracy of the final prediction. HistoROI was applied for quality control on a multi-organ dataset comprising 93 WSIs from the TCGA database, with manually annotated tissue segmentation. The average Dice score was 0.04 higher than the score obtained with HistoQC and was higher in 65 WSIs. However, during patch-level classification on the CRC-100k dataset, which contains colorectal cancer tissue, it misclassified over 23\% of patches, raising questions about its ability to generalize to other tissue types. The tool's code is publicly available at \url{https://github.com/abhijeetptl5/historoi} (last commit: May 23, 2025). Since the tool can identify both artifacts and specific tissue regions, such as lymphocyte-dense areas and adipose tissue, its multiclass predictions were converted into a binary representation for evaluation. The class explicitly labeled as an artifact, together with the background class, was considered non-tissue, whereas all remaining classes were classified as artifact-free tissue. The tool was not evaluated on dataset B because it does not allow specifying a zoom level as an input parameter for images saved as PNG files.

Building on HistoROI, \citet{patil2025semanticsegmentationbasedquality} developed a semantic segmentation algorithm that distinguishes artifact-free tissue, out-of-focus regions, tissue folds, and marker artifacts. The segmentation process utilizes four modules, each employing either a UNet \citep{journals/corr/RonnebergerFB15} or a UNet++ \citep{zhou2018unetplusplus} architecture with a ResNet18 or ResNet34 backbone. The authors evaluated the concordance between segmentation masks generated by HistoQC and those produced by the proposed tool across 11,666 images from the TCGA database, representing 28 organs. The Dice index was used to quantify agreement between the masks. In cases of low agreement between masks, pathologists more frequently judged the proposed solution as superior to HistoQC, although some segmentations from both methods were assessed as equally inadequate. The implementation of this method is publicly available at \url{https://github.com/abhijeetptl5/wsisegqc} (last commit: May 23, 2025). The blurring is quantified on a scale from 0 to 7, where 0 indicates no blurring and 7 indicates very high blurring. For evaluation, pixels were classified as out-of-focus if they received a non-zero value for that class. Additionally, the algorithm segments adipose tissue, which is considered artifact-free tissue during method evaluation.

SliDL \citep{berman_slidl_2023} is a Python library for pre- and post-processing of WSIs. It supports operations such as tissue and artifact segmentation, image tiling, and annotation management. SliDL uses a DenseNet architecture \citep{huang_densely_2017} for tissue and artifact segmentation. The output mask indicates background, tissue, and artifact regions. The model was trained on a dataset of 9,071 tiles derived from 393 annotations across 61 WSIs, scanned using various machines over different time periods and from multiple tissue types collected from two species. This diversity was intended to capture a wide range of WSI artifact variations, including pen marks, folded or torn tissue, slide bubbles, cracks, blurred or out-of-focus regions, uneven illumination, aberrant staining, and other marks. The artifact segmentation module was evaluated on a dataset comprising 36 HE-stained images from both public and private sources. In this study, the latest available version from the remote repository at \url{https://github.com/markowetzlab/slidl} (last commit: January 16, 2023) was used. However, the tool was excluded from the analysis because it failed to run, even though it was installed correctly according to the documentation. The error was caused by the model's trained weights being inaccessible. While the article included a section on common errors and solutions, it did not address this specific issue.

SlideInspect \citep{scotto_slideinspect_2026} is a recently developed framework designed for artifact segmentation and quality control. In addition to the pixel-level segmentation mask, it also takes into account information about the quality of tissue staining and provides feedback on the quality of the entire specimen in the form of a traffic light quality indicator (green, yellow, and red for slides of good, average, and poor quality, respectively). It also recommends actions to improve slide quality (re-scan, re-mount, re-cut, no action). For artifact segmentation, a K-Net architecture \citep{zhang_k-net_2021} with a Swin Transformer backbone \citep{liu_swin_2021} is used. Additionally, a machine-learning classifier is trained to produce the final recommended action for each slide based on stain quality and identified distortions. The tool also provides quantitative output, including the distribution of artifacts. The tool was evaluated on a large multicenter dataset comprising over 3,000 WSIs scanned across multiple scanners and four staining types. According to the authors, the framework demonstrates robustness across various tissue types, staining methods, and scanning platforms. However, this algorithm was excluded from the comparison due to the lack of a publicly available implementation. 

\citet{smit2021quality} developed a tool comparable to SlideInspect. It allows for multiclass artifact segmentation, including out-of-focus regions, tissue folds, ink, dust, markers, and air bubbles. The proposed method starts with tissue versus background segmentation using a resolution-agnostic CNN \citep{bandi_resolution-agnostic_2019}. Artifact segmentation is then performed with DeepLabV3+ \cite{chen_encoder-decoder_2018} employing an EfficientNet-B2 \cite{tan2020efficientnetrethinkingmodelscaling} encoder. Subsequently, the quality control module uses a decision tree to recommend one of four actions to improve slide quality: clean up, re-scan, re-cut, or no action. The training dataset consisted of 142 WSIs representing 9 tissue types and 8 staining types, scanned with 7 different scanners. Additionally, the algorithm was evaluated using a separate dataset of 30 unannotated images from an external data center. This algorithm achieved a Dice score of 0.91 on the internal test subset. Moreover, it demonstrated superior performance compared to HistoQC on the external dataset. However, information on improvements in segmentation quality metrics is not reported. The demo is publicly available on the Grand Challenge platform \citep{grandchallenge_wsi_quality}. However, due to missing model weights in the GitHub repository, technical limitations, and potential legal restrictions on uploading medical data, this tool is not suitable for use in medical institutions. Consequently, it was excluded from the comparison. 

\citet{schreiber_rapid_2024} developed a fast, simple thresholding-based method which is designed to eliminate artifacts such as pen markings and scanning artifacts. The algorithm begins by normalizing the RGB image to the range $\langle0,1\rangle$. Next, it performs element-wise multiplication on the outputs of two ReLU functions applied to the differences between the red and green channels and between the blue and green channels, respectively. Finally, Otsu thresholding is applied to segment the artifact-free tissue. The method requires no parameter tuning. The algorithm was evaluated on a dataset of 60 HE-stained duodenal biopsies scanned with different scanners, including 30 slides with artifacts such as marker traces and scanning artifacts. The method successfully segmented tissue in all cases and failed only once because it could not remove pink marker traces. This algorithm is relatively easy to implement, and the authors made the code publicly available. \url{https://gitlab.developers.cam.ac.uk/bas43/h_and_e_otsu_thresholding} (last commit: May 14, 2024). The tool assigns pixels to one of two categories: tissue without artifacts or background, and regions containing artifacts. In this study, tissue segmentation was performed at 10x magnification.

GrandQC \cite{wengGrandQCComprehensiveSolution2024a} is a state-of-the-art tool developed for tissue and multi-class artifact segmentation. It consists of two modules: the background versus tissue segmentation module and the artifact segmentation module. Both parts use EfficientNetB0 as the encoder and UNet++ as the decoder. The algorithm produces a pixel-wise segmentation mask that assigns each pixel one of the following classes: out-of-focus, pen marks, dark spots and foreign objects, air bubbles and edges (coverslip, broken glass), background, and artifact-free tissue. Tissue detection is performed at magnification 1x, while Artifact segmentation can be performed at three magnifications: 5x, 7.5x, and 10x. The lower the magnification, the shorter the computation time, with slightly lower detection quality. The multi-center dataset used to train the model consisted of 208 HE-stained WSIs for the tissue detection task and 420 WSIs for the artifact detection task. These two datasets partially overlapped. The tool was evaluated on a manually annotated dataset comprising 318 WSI images from a single pathology department, a subset of which was used during model evaluation in this study (see Section \ref{sec:datasets} for details). In addition, two pathologists evaluated the segmentation results on 600 WSIs from five medical centers. Their evaluation confirmed the algorithm's usefulness for identifying regions unsuitable for analysis. Furthermore, it has been shown that the method can be used for quality control within a given laboratory and for evaluating scanning systems. GrandQC was evaluated using patch-level quality segmentation metrics and compared with existing methods, demonstrating superior performance over HistoQC, HistoROI, and PathProfiler. However, only average Dice index results were reported, which prevents the analysis of cases with good and poor segmentation, and no statistical analysis of the obtained results was presented. Additionally, the segmentation masks provided by the authors for images from the TCGA database contain an error: the inference script does not account for padding at the bottom and right sides of the image, which means that if tissue is present there, that entire region is excluded from the segmentation mask and treated as background, leading to a loss of information. In this study, the publicly available version in a GitHub repository was used \url{https://github.com/cpath-ukk/grandqc} (last commit Dec 27, 2025). For artifact segmentation, the 10x magnification model was selected to provide the most accurate segmentation results.

In the next work, the GrandQC tool was used to perform quality control on 544 randomly selected WSIs to assess its usefulness in real laboratory settings \citep{borrecho_grandqc_2026}. The authors enhanced the original method by developing additional scripts to quantify the number of pixels associated with each artifact type and to generate an Excel file for statistical analysis, which they refer to as GrandQC+. The results demonstrated that GrandQC+ enables quick data quality assessment and facilitates the identification of low-quality slides. However, the developed scripts are not publicly available.

HistoKit \citep{bioimaging26} is a publicly available tool for WSI processing. It supports tissue and artifact segmentation, patch extraction, data augmentation, and stain normalization. For tissue detection, HistoKit employs a thresholding-based segmentation method \citep{bioimaging25} that utilizes the GaMRed algorithm \citep{marczyk2020gamred}.  During tissue detection, the algorithm also removes low-chroma components and marker traces. For artifact detection, HistoKit combines the GrandQC artifact segmentation module with an improved inference algorithm. This enhancement introduces patch overlaps to minimize discordances at patch borders, aiming to improve segmentation quality, particularly in smaller or less convex tissue regions. The predictions in overlapping patch regions can be aggregated using the \textit{Constant} or \textit{Gaussian} modes. The first approach calculates the arithmetic mean of the confidence scores for each artifact class and assigns each pixel in the final mask to the class with the highest mean confidence score. Alternatively, a weighted average can be applied, with the weights determined using a Gaussian kernel with a specified standard deviation. HistoKit also provides confidence maps for the entire image for each artifact class. This feature allows users to exclude specific artifact classes from subsequent analyses. It also generates a detailed report of the artifacts found in each image. Although the tool has not yet been evaluated for its impact on segmentation quality using standard segmentation metrics, the reported visual results demonstrate a reduction in artifacts resulting from patch processing. In this study, version 1.0.1 of HistoKit, available at  (\url{https://github.com/ZAEDPolSl/HistoKit}), was used. Tissue segmentation can be performed at any magnification level, but higher magnification increases the segmentation quality. Both tissue detection and artifact segmentation were conducted at 10x magnification. For algorithm comparison, the constant aggregation mode with 75\% patch overlap was used to combine predictions in overlapping regions.

GuardQC \citep{wei_tissue-preserving_2026} is a tool that incorporates the GrandQC artifact segmentation module as its core component. The module generates confidence score maps and connected components, which serve as proposal inputs and are subsequently processed by a dual-stream ResNet18 model. This approach integrates local texture information with tissue context following dynamic multiresolution alignment. Then, the classifier head outputs the probability of artifact presence, enabling the exclusion of regions containing artifact-free tissue prior to any pixel-level refinement. To enhance specificity, iterative hard negative mining \citep{7780458} is applied to prevent misclassification of dark cell nuclei as artifacts. Patches containing tissue likely to confuse the algorithm, such as those with cell nuclei and an artifact probability exceeding the 55\% threshold, are combined with easily classifiable, artifact-free patches. This set is used for fine-tuning the model over 54 iterations. Finally, candidates with an artifact probability exceeding the threshold undergo local refinement via adaptive thresholding. For network evaluation, a subset of the GrandQC test set was used, consisting of 23 whole slide images (WSIs) from the BEETLE multicentric breast cancer segmentation dataset \citep{lems_2025_16812932}. GuardQC reduced the number of artifact-free tissue pixels classified as artifacts in comparison with GrandQC. However, the tool's source code is not publicly available, so it was excluded from the comparison.

AIRAQc \citep{gautam2025airaqc} is another deep-learning-based tool for quality control. The algorithm uses a multiple-magnification transformer architecture for tissue and artifact segmentation. It assumes that some types of artifacts, such as marker traces, can be detected at lower magnification than artifacts in out-of-focus regions. It processes images at three different magnifications (2.5x, 5x, and 10x) using a transformer architecture with shared weights across magnification levels, followed by cross-magnification feature fusion. Finally, a lightweight multilayer perceptron (MLP) decoder is employed to produce the final segmentation mask. This algorithm was trained on a large private dataset comprising 40,000 WSIs from various medical data centers. The training dataset included two staining types: HE and IHC. The tool was evaluated on 50 WSIs from the TCGA database, including 25 artifact-free slides. A comparison of the averaged Dice index for various artifact classes and artifact-free tissue segmentation demonstrated that AIRAQc outperformed HistoQC, HistoROI, and GrandQC. However, the results were not statistically analyzed, and confidence intervals were not provided. Moreover, the tool's code is not publicly available.

DiffusionQC \citep{wang2026diffusionqcartifactdetectionhistopathology} uses a distinct approach, treating quality control as an outlier detection problem. The tool has two architectural variants based on the PixCell-1024 latent diffusion model \citep{yellapragada2025pixcellgenerativefoundationmodel}. The baseline variant learns the distribution of artifact-free images to detect artifacts as outliers. The enhanced variant additionally uses a contrastive learning component to explicitly push artifact images away from clean images in the latent space. During training, precise artifact segmentations are not needed. Instead, images were divided into patches and classified as either containing artifacts or being artifact-free. A smooth segmentation mask is obtained by applying a straightforward post-processing pipeline that includes blurring the spatial error heatmap, followed by thresholding and morphological operations. However, the parameters used in this step should be selected by the user according to the specific dataset and staining method. To train the model, 50 HE-stained slides from the TCGA database were used, which were made available by the authors of the ARIAQc tool. The enhanced version of DiffusionQC outperformed GrandQC on the test subset in sensitivity, precision, and F1 scores. However, the training and test sets were obtained by splitting the same dataset. Furthermore, no confidence intervals or standard deviations were provided, making it impossible to determine whether the method achieves high-quality segmentation for all images. The tool's source code is not available, so it has been excluded from this comparison.

\section{Datasets}
\label{sec:datasets}

\begin{figure*}[h!]
    \centering
     
    \includegraphics[
    width=0.75\linewidth,
    trim=1cm 0cm 0cm 3.5cm,
    clip
]{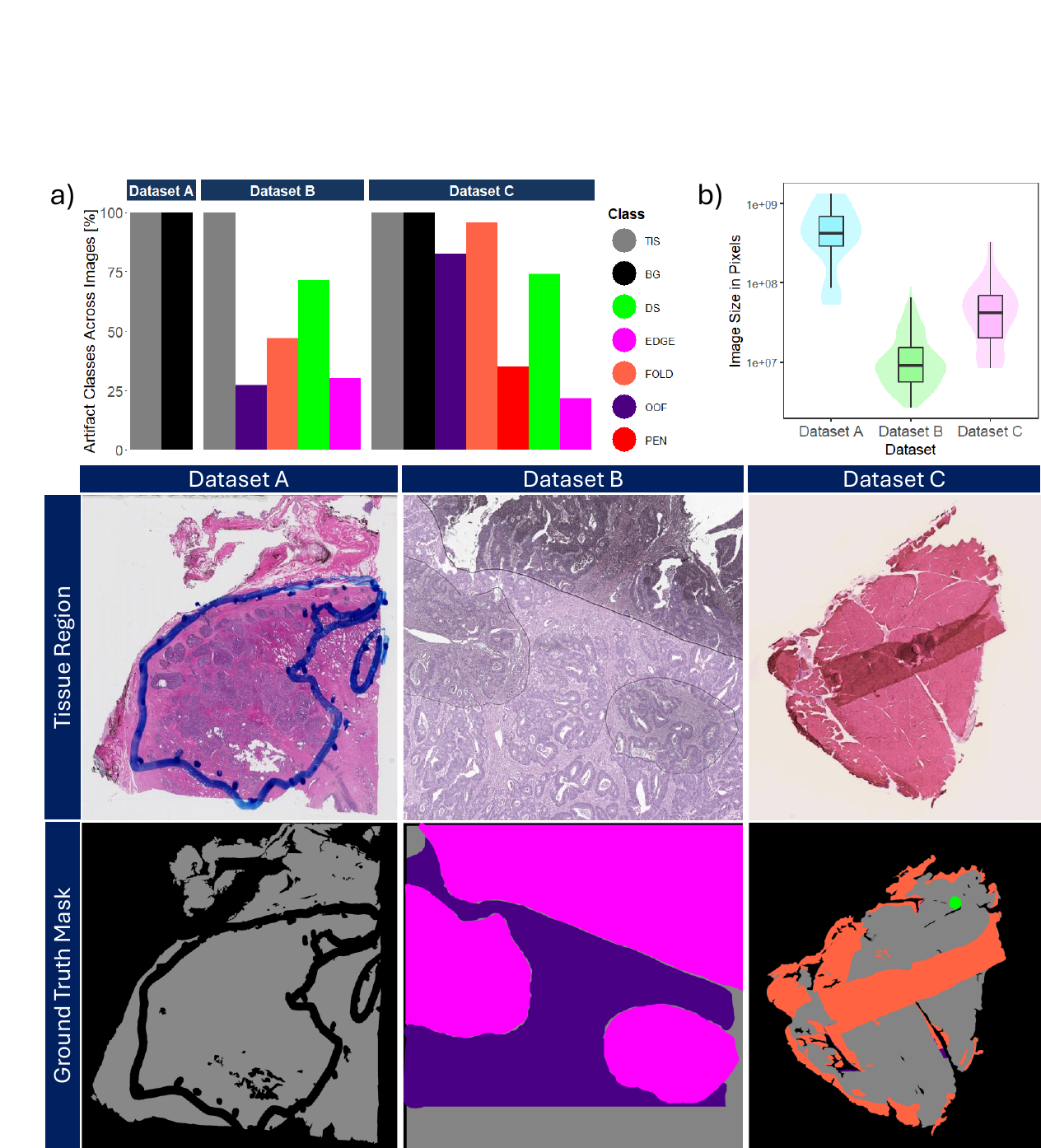}
    \caption{Overview of the datasets used in the study. (a) Percentage of images containing each selected artifact type across the datasets. In Dataset A, the background class includes pixels corresponding to both the image background and other artifact types, such as markers, tissue folds, and foreign objects. (b) Plot showing the size of the annotated tissue regions in pixels for each dataset. The y-axis is shown on a logarithmic scale. (c) Representative images from each dataset and their corresponding ground-truth annotations.}
   \label{fig:datasets}
\end{figure*}

Tissue quality control is typically performed before patch extraction. This process requires comprehensive information about the entire tissue area, as artifacts such as air bubbles or blurring may extend across multiple patches. Manual labeling of artifacts is highly time-consuming, thereby limiting the number of datasets available for benchmarking. Consequently, many datasets provide annotations only for separated patches of a fixed size and magnification, lacking information about the broader tissue context. In addition, datasets often contain only a single label indicating the artifact class assigned to the entire patch, without a precise segmentation mask \citep{wang2020focuslitenn, kanwal_2024_10809442}. Different algorithms often employ varying patch sizes, and significantly increasing patch size can introduce interpolation artifacts, potentially biasing results due to data availability rather than the algorithm's performance \citep{Panda2024ImageUpscaling}. Furthermore, evaluating algorithms solely at the patch level does not reveal their performance on tissue specimens, where distortions often occur at patch intersections. Without procedures to eliminate these disturbances, segmentation outcomes in these regions may be unreliable. To determine whether algorithms can address these challenges, this study employed three image datasets that include larger tissue regions: two publicly available datasets and one newly introduced dataset, thereby more closely simulating real-world laboratory conditions. The datasets vary in both the size of the tissue regions and the types of artifacts present. Figure \ref{fig:datasets} shows sample images along with segmentation masks from each dataset. Table \ref{tab:datasets} summarizes the datasets used to assess artifact segmentation methods. Based on the artifact classification proposed by \citet{wengGrandQCComprehensiveSolution2024a}, the following artifact classes were considered in this study:

\begin{itemize}
\item \textbf{Tissue (TIS)} -- tissue regions without artifacts.

\item \textbf{Background (BG)} -- pixels corresponding to the slide background. For Dataset A, in the binary classification task, this class also includes pixels corresponding to tissue and other artifact types.

\item \textbf{Dark Spot (DS)} -- dark regions caused by foreign objects, such as dust particles and hair, or by non-uniform illumination.

\item \textbf{Edge (EDGE)} -- artifacts including coverslip edges, air bubbles, and cracked glass.

\item \textbf{Fold (FOLD)} -- regions containing tissue folds.

\item \textbf{Out-of-Focus (OOF)} -- image regions affected by blurring caused by the scanner movement being too fast or other factors.

\item \textbf{Pen (PEN)} -- regions containing pen or marker traces of various colors.

\end{itemize}

\begin{table*}[t!]
\centering
\caption{Summary of datasets used to benchmark artifact segmentation algorithms.}
\label{tab:datasets}
\resizebox{\textwidth}{!}{%
\begin{tabular}{@{}lccll@{}}
\cmidrule(r){1-4}
\multicolumn{1}{c}{}                            & \textbf{Dataset A }           & \textbf{Dataset B}                 & \multicolumn{1}{c}{\textbf{Dataset C}}                          &  \\ \cmidrule(r){1-4}
                                                & \multicolumn{1}{l}{} & \multicolumn{1}{l}{}      &                                                        &  \\
\multicolumn{1}{c}{\textbf{Number of Annotated Regions}} & 13                   & 567                       & \multicolumn{1}{c}{23}                                 &  \\
                                                &                      & \multicolumn{1}{l}{}      &                                                        &  \\
\multicolumn{1}{c}{\textbf{Source}}                      & TCGA, CAMELYON16     & Medical Center in Austria & \multicolumn{1}{c}{Compass NMD, TCGA}                  &  \\
                                                &                      & \multicolumn{1}{l}{}      &                                                        &  \\
\multicolumn{1}{c}{\textbf{Formats}}                     & SVS / TIF            & PNG                       & \multicolumn{1}{c}{SVS}                                &  \\
                                                &                      & \multicolumn{1}{l}{}      &                                                        &  \\
\multicolumn{1}{c}{\textbf{Organs}} &
  \begin{tabular}[c]{@{}c@{}}Breast, Head and Neck, Nervous system, \\ Prostate, Small Intestine\end{tabular} &
  Breast, Colon, Kidney, Prostate &
  \multicolumn{1}{c}{\begin{tabular}[c]{@{}c@{}}Brain, Head and Neck, Heart,  \\ Kidney, Muscle, Uterus\end{tabular}} &
   \\
                                                &                      & \multicolumn{1}{l}{}      &                                                        &  \\
\multicolumn{1}{c}{\textbf{Preservation Method}}         & FFPE, Unknown        & FFPE                      & \multicolumn{1}{c}{FFPE, Fresh Frozen Tissue, Unknown} &  \\
                                                & \multicolumn{1}{l}{} &                           &                                                        &  \\
\multicolumn{1}{c}{\textbf{Annotated Classes}} &
  \begin{tabular}[c]{@{}c@{}}Artifact-Free Tissue, \\ Background and Artifacts\end{tabular} &
  \begin{tabular}[c]{@{}c@{}}Air Bubbles and Edges, Artifact-free Tissue, \\ Dark spots and Foreign Objects, \\ Out-of-Focus Regions, Tissue Folds\end{tabular} &
  \multicolumn{1}{c}{\begin{tabular}[c]{@{}c@{}}Air Bubbles and Edges, Artifact-Free Tissue, \\ Background, Dark Spots and Foreign Objects, \\ Out-of-Focus Regions, Pen, Tissue Folds\end{tabular}} &
   \\
                                                & \multicolumn{1}{l}{} &                           &                                                        &  \\
\multicolumn{1}{c}{\begin{tabular}[c]{@{}c@{}}\textbf{Mean Image Size} \\\textbf{ with 95\% CI }\textbf{($10^6$ pixels)}\end{tabular}} &
  652.98 (368.18--937.78) &
  13.10 (12.14--14.07) &
  \multicolumn{1}{c}{56.24 (27.86--84.62)} &
   \\
                                                & \multicolumn{1}{l}{} & \multicolumn{1}{l}{}      &                                                        &  \\ \cmidrule(r){1-4}
\end{tabular}%
}
\end{table*}

\subsection{Dataset A} 
This dataset is a publicly available subset of the test set provided by \citep{Berman2023SliDL}. It contains WSIs from the TCGA database and the CAMELYON16 challenge. Manual annotations were made publicly available with the tutorial repository provided with the SliDL tool \url{https://github.com/markowetzlab/slidl-tutorial} (last commit: Feb 14, 2022). Each slide and mask has been carefully reviewed. The masks contain binary annotations distinguishing artifact-free tissue from artifacts and background regions. Artifacts identified on the slides include tissue folds, marker traces of various colors, air bubbles, and dust. Although blurred areas are present on some tissue sections, these regions were not annotated in the masks. Consequently, in analyses employing methods that support multi-class artifact segmentation, blurred areas were considered artifact-free tissue.

\subsection{Dataset B} 
The dataset analyzed in this study is a subset of the training set provided by the authors of the GrandQC \citep{wengGrandQCComprehensiveSolution2024a}. The original dataset included manual annotations of tissue regions extracted from 318 WSIs, representing 80 cases, from a single pathology department. The annotated tissue types included kidney, colon, prostate, and breast. An experienced pathologist performed the annotations using QuPath \citep{Bankhead2017QuPath}. Various artifacts, including tissue folds, out-of-focus regions, bubbles (simulated with glue droplets), and foreign objects (simulated with threads), were introduced by an experienced pathologist. Although the authors reference a dataset containing marker traces, such slides are not present in the publicly available data. The dataset is divided into patches, each with information regarding its location on the slide. Therefore, it was possible to reassemble the larger tissue regions. To evaluate tool performance on larger images, only regions containing at least three patches in both the vertical and horizontal directions were retained. Additionally, regions containing more than 30\% background pixels or those where less than 10\% artifact-free tissue was present were excluded. This filtering process resulted in 574 images. The dataset also includes information about pixels assigned to the background class, but these labels are highly inconsistent. In some cases, holes in the tissue are labeled as tissue, while in others they are marked as background in the annotation mask. Consequently, pixels assigned to the background class on the annotation mask were excluded from the analysis, and pixels assigned to the background class on the prediction mask were consistently replaced with the tissue class.
 
\subsection{Dataset C}
This dataset comprises 23 manually annotated tissue regions from 20 WSIs, accompanied by detailed segmentation masks. The tissue slides originate from two sources: publicly available WSIs from TCGA database representing six different organs, and slides from the Compass NMD project \cite{compass_nmd_2023} containing muscle tissue. Both complete SVS files and tissue regions saved as single-layered TIF files at 10x magnification are provided, along with metadata specifying the location of each tissue region on the slide. Annotation was performed by a doctoral student in computational pathology (JM) using ImageJ \citep{Schneider2012ImageJ}. The Polygon and Manual Selections tools were used to ensure precise, detailed annotations. Particular care was taken when annotating regions affected by blurring, as these areas are challenging to delineate manually. Further details about the dataset are available in the supplementary materials. 

\section{Evaluation}

In the first task, the methods were compared based on their ability to distinguish artifact-free tissue from both the background and regions containing artifacts. In the second task, GrandQC, HistoKit, and \citep{patil2025semanticsegmentationbasedquality} were evaluated for their ability to identify specific artifact classes. 

Artifact-free tissue segmentation can be formulated as a binary classification problem at the pixel level, in which tissue is the positive class and artifacts and background are the negative class. In this context, undetected artifacts were considered false positives (FP), undetected tissue regions were false negatives (FN), correctly classified tissue pixels were true positives (TP), and correctly classified background or artifact pixels were true negatives (TN). To compare artifact segmentation models, three performance metrics were employed: Dice coefficient, pixel sensitivity, and pixel False Discovery Rate (FDR). These metrics are defined as follows:

\begin{equation}
Dice = \frac{2 \cdot TP}{(TP+FP+FN)}
\end{equation}

\begin{equation}
Sensitivity = \frac{TP}{TP+FN}
\end{equation}

\begin{equation}
FDR = \frac{FP}{(FP + TP)}
\end{equation}

All metrics range from 0 to 1. Higher Dice index and recall values correspond to more accurate artifact-free tissue segmentation. In contrast, lower FDR values indicate that more regions containing artifacts were segmented as artifact-free tissue.

To compare algorithms for multi-class artifact segmentation, the Dice score was calculated separately for each class. Pixels associated with a specific artifact were treated as the positive class, while all other pixels were considered the negative class. Consequently, pixels classified as both tissue and an incorrect artifact class were identified as false positives. The Dice index was calculated only for images containing a given type of artifact, so the number of images may vary for each artifact type. In addition, an analysis of misclassified pixels in a given dataset was conducted. Supplementary materials include examples of unlabeled WSIs from the Compass NMD project, processed with algorithms that achieved the best results on annotated datasets. 

The Friedman test \citep{Friedman01121937} was used to reveal significant differences among the methods for each dataset and evaluation metric. Pairwise comparisons were subsequently performed using the Wilcoxon signed-rank test \citep{nonparam_statistics}, with the Benjamini-Hochberg correction \citep{Chen2017MultipleComparisons} applied to account for multiple comparisons. The matched-pairs rank-biserial correlation coefficient \citep{FielPeres2026} was used to quantify the effect size. 

\section{Results}
\subsection{Segmentation of Artifact-Free Tissue}
Figure 3 shows sample images from each dataset processed using the available artifact segmentation methods, while Figure 4 presents the distributions of the Dice Index, sensitivity, and FDR for each method across all datasets. HistoQC did not process three images in dataset A because it could not obtain the image magnification (Supplementary Materials, Section S3). Therefore, for this method, the analysis was performed only on the processed images. Detailed results of the statistical analysis are provided in Supplementary Tables S1-S3. 

\begin{sidewaysfigure*}
    \centering
    \includegraphics[
        width=\textheight,
        trim=0cm 1.2cm 0.3cm 0cm,
        clip
    ]{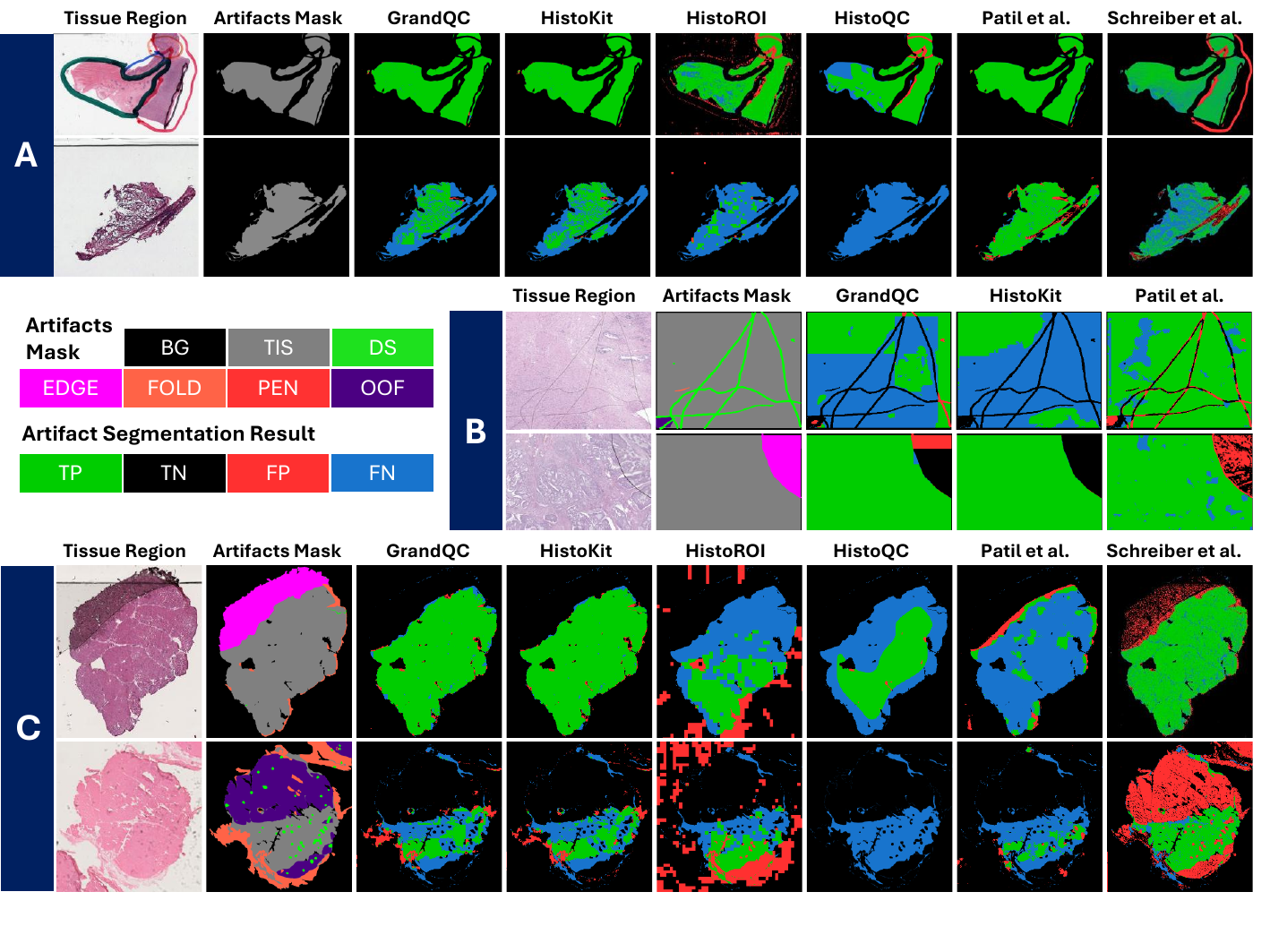}
    
    \caption{Exemplary segmentation results for each method across all datasets. 
    The dataset name (A, B, or C) is indicated in the leftmost panel. For each example, 
    the original tissue image and the corresponding ground-truth mask are shown, 
    with colors representing the annotated artifact classes. The segmentation 
    results obtained with each method are visualized using classification outcome 
    maps: true positive (TP) pixels are shown in green, true negative (TN) pixels 
    in black, false positive (FP) pixels in red, and false negative (FN) pixels 
    in blue.}
    
    \label{fig:cases_binary}
\end{sidewaysfigure*}

\begin{figure*}[h!]
    \centering
    \includegraphics[width=1\linewidth]{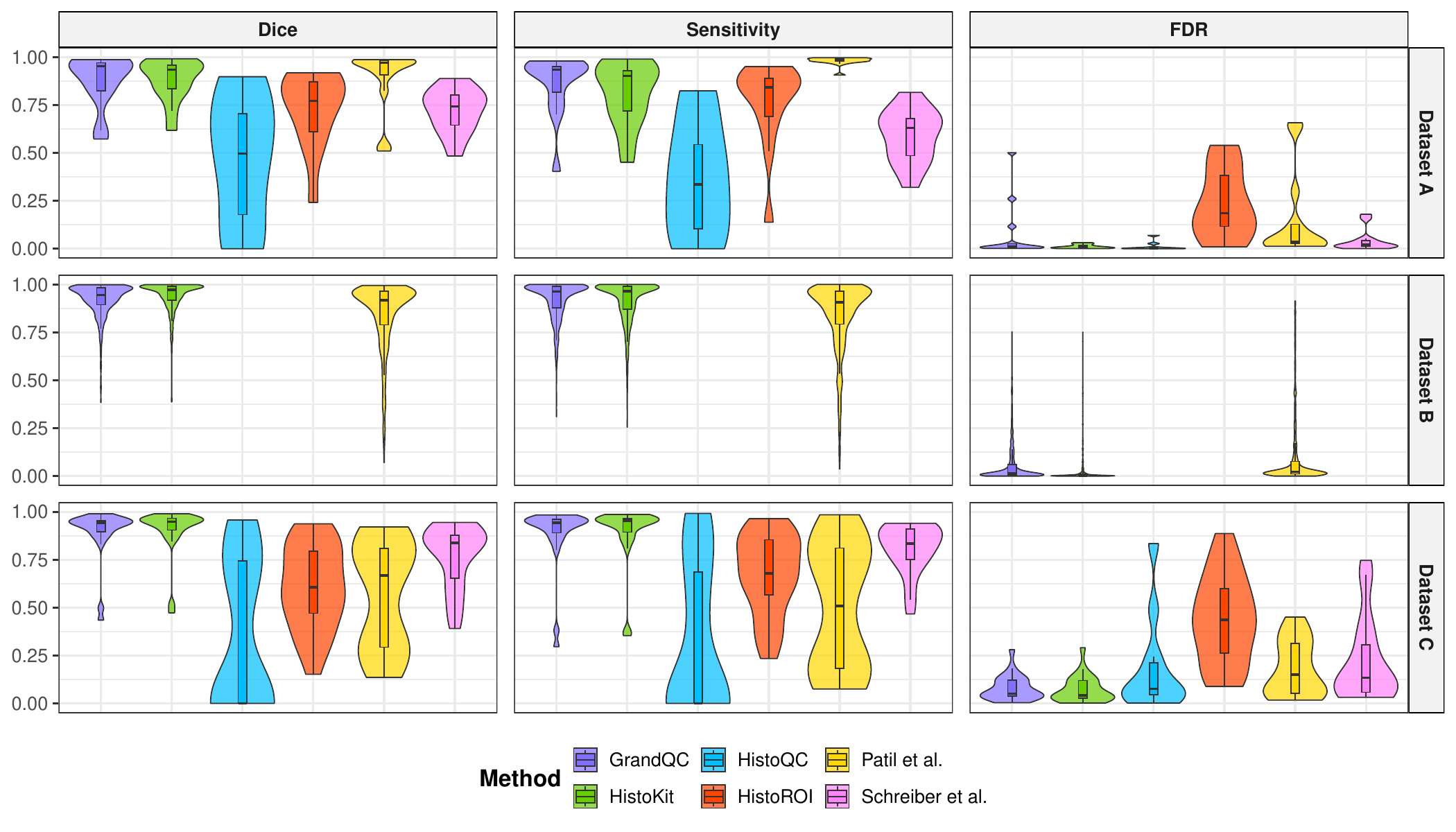}
    \caption{Violin plots showing the distributions of the Dice index, sensitivity, and FDR across color-coded methods and datasets.}
    \label{fig:results_violin}
\end{figure*}

HistoKit and GrandQC performed strongly across all datasets. Statistical analysis showed that HistoKit was significantly superior to GrandQC for Dataset B in Dice and FDR, and for Dataset C across all metrics. Moreover, the effect size indicated that HistoKit had an advantage in Dice and FDR, but the difference was not statistically significant. Because GrandQC did not overlap patches during prediction, sharply defined patch edges frequently appeared as discordance, leading to less precise segmentation. Additionally, square-shaped regions that appear randomly on artifact-free tissue were often misclassified as artifacts, or artifacts were misclassified as normal tissue. This limitation was not observed with HistoKit. 

For the algorithm developed by \citet{patil2025semanticsegmentationbasedquality}, results varied across datasets. This method obtained higher sensitivity scores than HistoKit and GrandQC for dataset A, but it performed worse in terms of FDR for this dataset. For Dataset B, it achieved good-quality segmentations for most images. However, it exhibited higher FDR values than HistoKit and GrandQC across all metrics. Moreover, this method outperformed HistoQC and \citet{schreiber_rapid_2024} in Dice and sensitivity. It also significantly improved sensitivity over its predecessor, HistoROI. For Dataset C, \citet{patil2025semanticsegmentationbasedquality} was significantly inferior to HistoKit and GrandQC across all metrics, and to \citet{schreiber_rapid_2024} in sensitivity and FDR. No significant differences were observed between this tool and HistoQC for this dataset. Compared with HistoROI, it reduced FDR scores, while no statistically significant differences were observed for other measures. In most cases, the tool detected excessively blurred regions but could not detect artifacts such as air bubbles or foreign objects.

HistoROI achieved better Dice scores and sensitivity than HistoQC for datasets A and C. Nevertheless, HistoQC outperformed it in terms of FDR. The scores obtained by HistoROI are attributed to its tendency to detect tissue in background areas or regions containing artifacts. In contrast, the lower number of false-positive pixels predicted by HistoQC results from its frequent failure to detect tissue, which may exclude WSIs from further analysis and unnecessarily reduce dataset size. Furthermore, on slides where tissue was segmented, the resulting masks often exhibited significantly smoothed edges, which also influenced mask size.

The method proposed by \citet{schreiber_rapid_2024} achieved results on Dataset A comparable to those of HistoROI and HistoQC, while demonstrating lower FDR scores than HistoROI. On Dataset C, it outperformed HistoQC, HistoROI, and \citet{patil2025semanticsegmentationbasedquality} in Dice and sensitivity, while achieving statistically significant improvements in FDR over HistoROI. While the approach effectively segmented tissue areas in most cases, it could not eliminate artifacts such as red marker traces, blurring, or air bubbles. The thresholded tissue mask contained numerous small objects and holes, which post-processing techniques such as morphological operations could address.

In addition, the results show that deep learning algorithms trained on large, diverse datasets generally achieve better results than methods that rely on thresholding and image features such as brightness and contrast, or algorithms trained on smaller, more homogeneous datasets.

\subsection{Multiclass Artifact Segmentation}

\begin{figure*}[h!]
    \centering
    \includegraphics[
        angle=0,
        height=0.9\textwidth,
        trim=0cm 0cm 2.3cm 0cm,
        clip
    ]{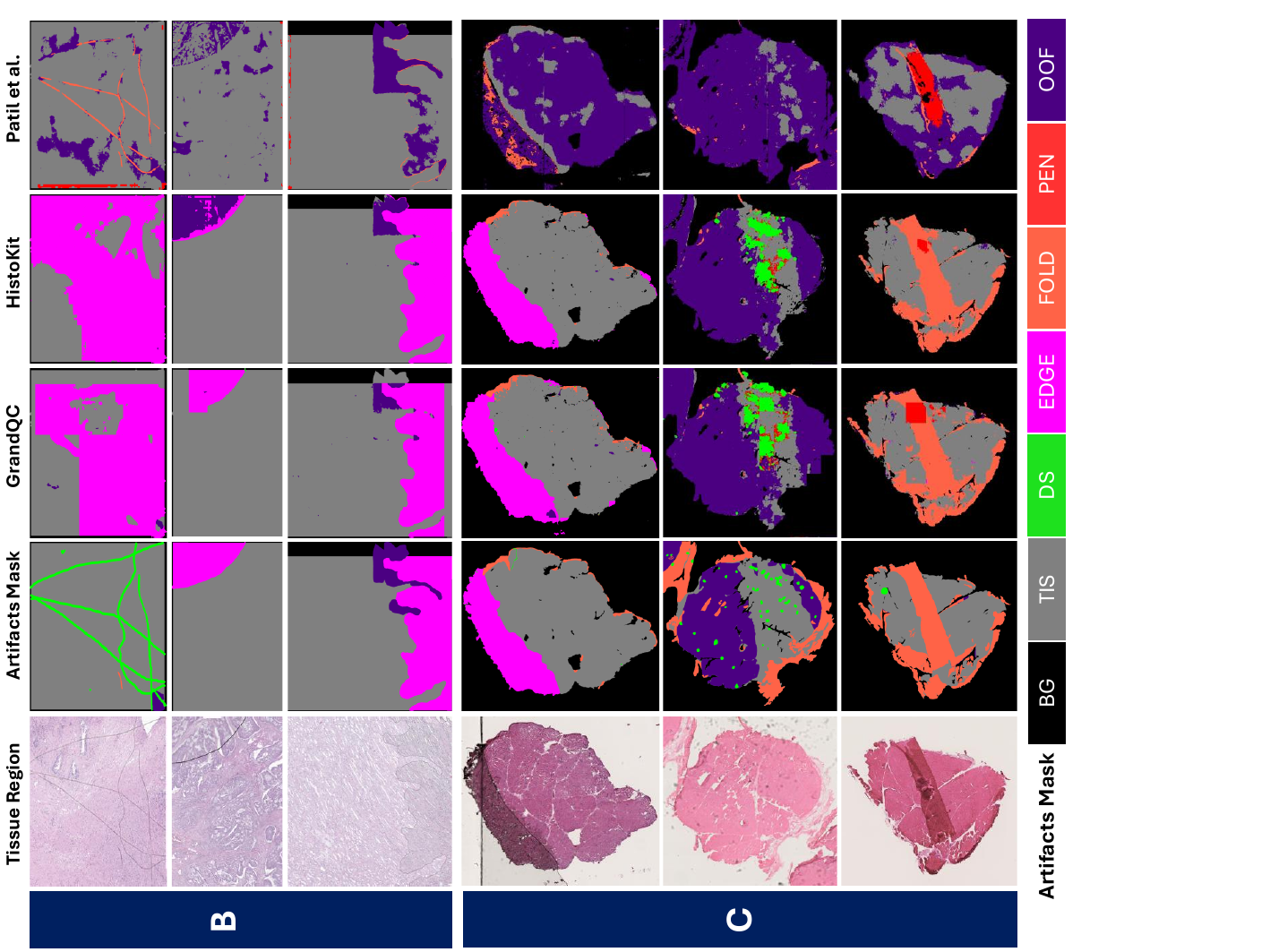}
    \caption{Tissue regions from Datasets B and C with the corresponding ground-truth multiclass artifact annotation masks and segmentation results for selected images from each dataset. Colors indicate different artifact classes: background (BG), artifact-free tissue (TIS), dark spots and foreign objects (DS), edges and air bubbles (EDGE), tissue folds (FOLD), pen markings (PEN), and out-of-focus regions (OOF).}
    \label{fig:multiclass}
\end{figure*}

\begin{figure}[h!]
    \centering
    \includegraphics[width=\linewidth]{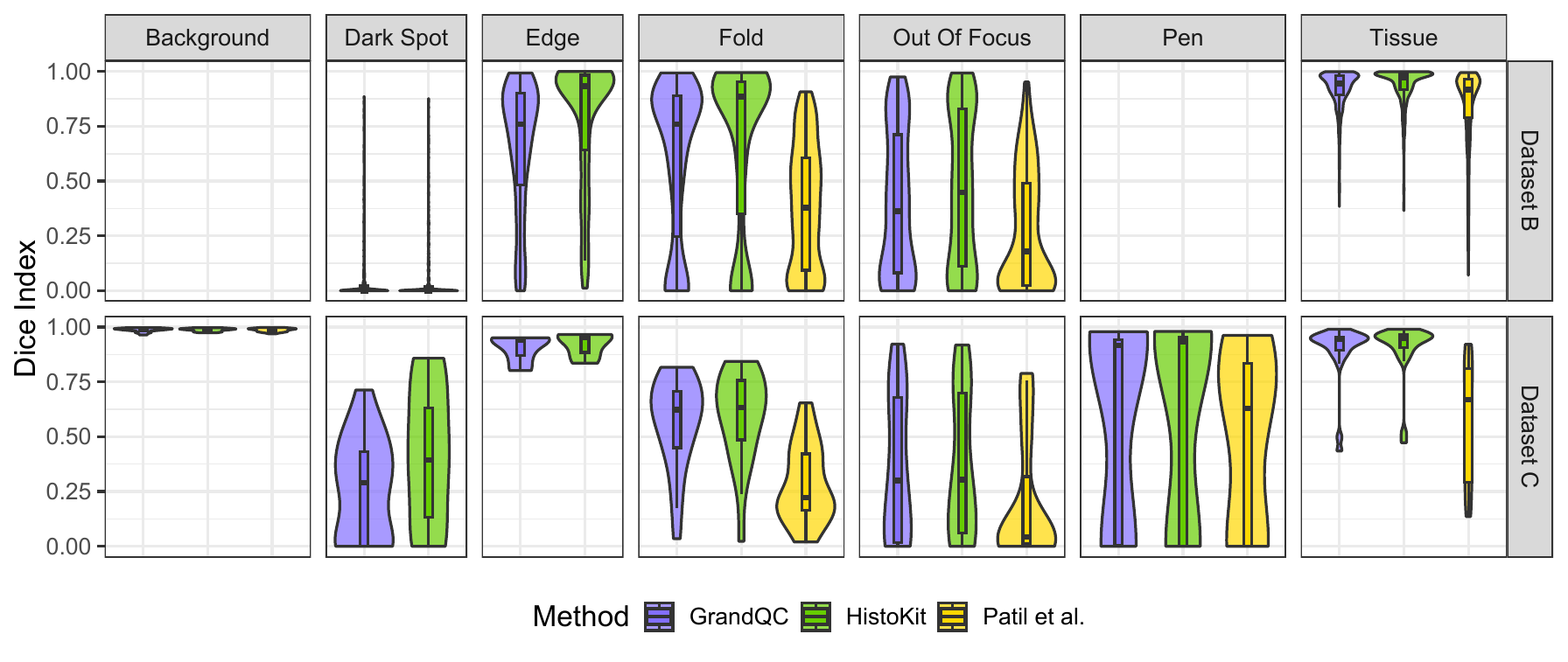}
    \caption{Violin plots showing the distributions of Dice Index values between methods across datasets, stratified by individual artifact classes. In Dataset B, the Background and Pen classes are unavailable. Since the method developed by Patil et al. does not support detecting artifacts in the Edge and Dark Spot classes, it is not shown in the corresponding plots.}
    \label{fig:violin_multiclass}
\end{figure}

\begin{figure}[h!]
    \centering
    \includegraphics[width=1\linewidth]{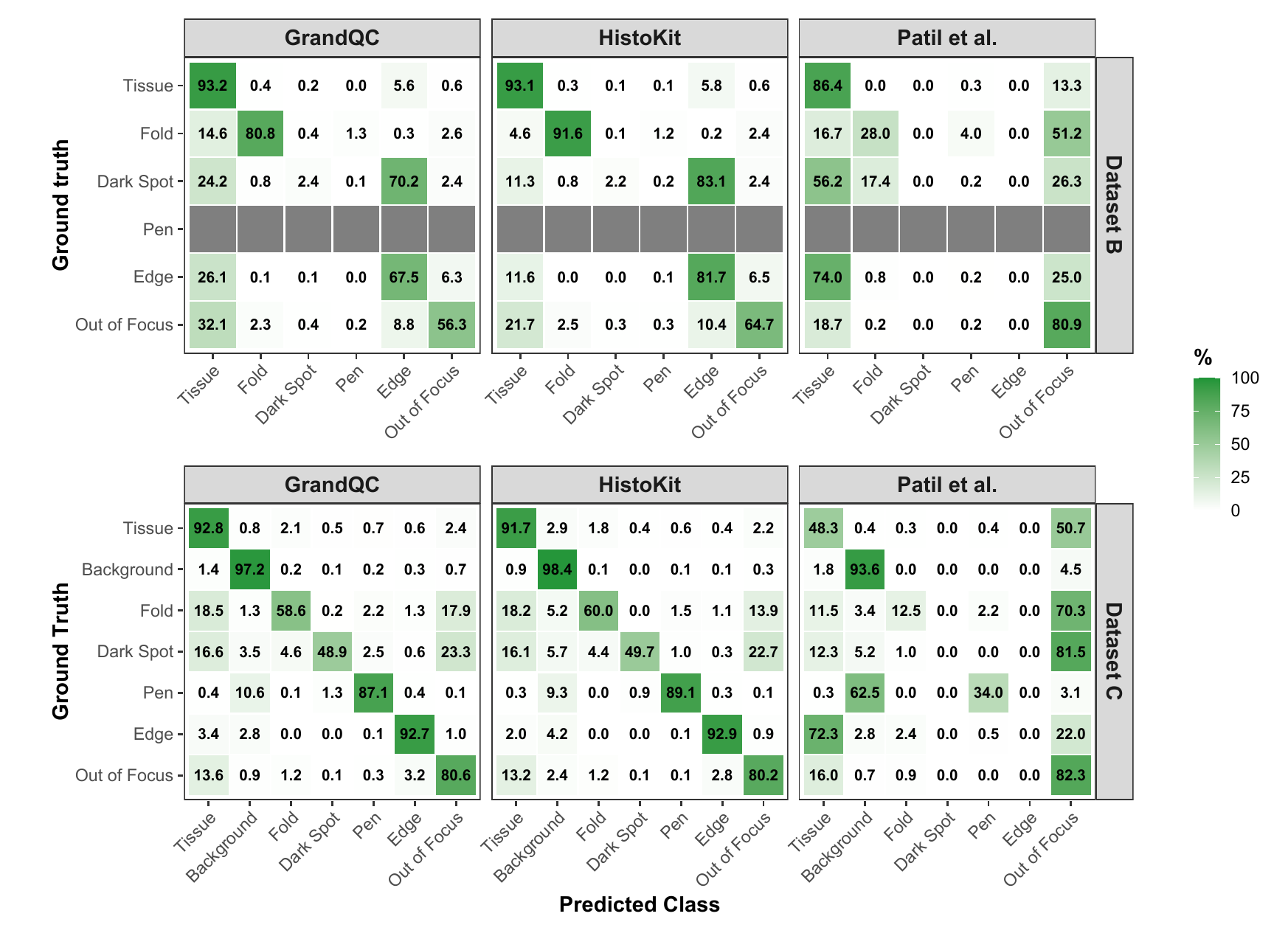}
    \caption{Heatmaps showing the percentage of pixels classified into each artifact class for datasets B and C. Although the Pen class is not present in dataset B, some algorithms incorrectly classify pixels as belonging to this class. Therefore, we include it in the heatmap. Gray indicates that a given artifact class is absent from the ground-truth image.
}
    \label{fig:heatmaps}
\end{figure}

Figure \ref{fig:multiclass} shows the results of artifact segmentation for exemplary images from Datasets B and C using methods that enable multiclass artifact segmentation: GrandQC, HistoKit, and \citet{patil2025semanticsegmentationbasedquality}, while Figure \ref{fig:violin_multiclass} shows the distributions of Dice scores for each artifact type. Because the Background and Pen classes are not present in Dataset B, results for these classes are not reported. The \citet{patil2025semanticsegmentationbasedquality} method does not detect artifacts in the Edge and Dark classes. Consequently, the corresponding Dice values are not shown on the plot. In addition, Figure \ref{fig:heatmaps} shows heatmaps containing information on the percentage of pixels classified into correct and incorrect artifact classes. In this case, the Edge and Dark Spot classes for method \citet{patil2025semanticsegmentationbasedquality} were not omitted in order to determine whether the method assigns these pixels to a tissue class or labels them as some other artifact class.

Again,  statistical analysis was performed to compare the results obtained from the algorithms (Supplementary Table S4). When only two methods distinguished a particular artifact type, the Friedman test was not conducted. The algorithms showed no significant differences in background segmentation. However, for other classes, GrandQC and HistoKit achieved superior results compared to \citet{patil2025semanticsegmentationbasedquality} on both datasets. Furthermore, HistoKit outperformed GrandQC in multiclass artifact segmentation, achieving better or similar results for each class. HistoKit produced a statistically significant improvement in Dice scores for the Edge class in Dataset B. In Dataset C, statistical significance was not achieved, likely due to the limited number of images. Nevertheless, the effect size indicates a meaningful improvement over GrandQC. HistoKit also enhanced segmentation of tissue folds in both datasets, the Darkspot class in Dataset C, and the out-of-focus regions in Dataset B. Notably, \citet{patil2025semanticsegmentationbasedquality} performs poorly at classifying pixels belonging to Fold and Out-of-Focus classes, particularly in Dataset C.

For Dataset B, both GrandQC and HistoKit correctly classified most tissue pixels, achieving comparable results. In both algorithms, tissue pixels were most often misclassified as Edge. Notably, HistoKit significantly reduced the misclassification of Fold class pixels as Edge class, as indicated by the Dice scores. This improvement was also evident for Edge class in dataset B, where these artifacts were more prevalent than in dataset C. Moreover, for dataset C, the number of pixels correctly assigned to the Pen class also increased. For Dark Spot, GrandQC and HistoKit produced similar results. In both datasets, HistoKit reduced the number of artifact pixels that GrandQC misclassified as tissue.

Conversely, in both datasets, Patil et al. classified many artifacts and pixels belonging to artifact-free tissue as the Out-of-Focus class. Also, many artifacts in the Fold class were treated as Tissue. This method was not designed to segment regions containing artifacts belonging to the Edge and Dark Spot classes. Some artifacts of other types are classified as Out-of-Focus.

Supplementary Figures S1-S2 present a comparison of sample segmentations from GrandQC and HistoKit for two images from the Compass NMD project without expert annotations. Isolated predictions resembling patch boundaries are common, and the HistoKit strategy helps reduce these distortions. The results again demonstrated the significant advantage of methods trained on a diverse training dataset and showed that using overlap between patches during prediction not only allows more accurate segmentation of artifact-free tissue versus artifacts, but also reduces errors when classifying pixels across artifact classes. 

\subsection{Computational time comparison}

Table \ref{tab:time} presents the mean and confidence intervals for 10 segmentation time measurements of a single image from set C (file size equal to 0.98 GB), using the selected algorithms. The experiments were conducted on a system equipped with an Intel\textsuperscript{\textregistered} Xeon\textsuperscript{\textregistered} Gold 6542Y processor, 504~GB of RAM, and one NVIDIA RTX 6000 Ada GPU. For HistoKit, during the artifact segmentation step, the batch size was set to 32 and 6 workers were used for patch extraction. The HistoROI method demonstrated the fastest performance, while \cite{schreiber_rapid_2024}, \cite{patil2025semanticsegmentationbasedquality}, and HistoQC exhibited slightly slower processing times. Tissue detection with GrandQC is rapid because it is performed at 1x magnification. In contrast, HistoKit conducts tissue segmentation at 10x magnification to ensure accuracy, leading to a substantial increase in processing time. The multithreaded implementation of prediction in HistoKit is nearly twice as fast as the original GrandQC algorithm when no overlap is used. Increasing the overlap raises the number of processed patches, which extends processing times but enhances the algorithm’s accuracy (see Section \ref{sec:modification}).

\begin{table}[h!]
\centering
\caption{Comparison of segmentation times for the selected algorithms. For GrandQC and HistoKit, the processing times of individual stages are also reported. For HistoKit, the relationship between processing time and patch overlap is additionally presented. The mean processing time was calculated from 10 segmentation runs performed on an image from dataset C.}
\label{tab:time}

\resizebox{0.6\columnwidth}{!}{%
\begin{tabular}{lcll}
\hline
\textbf{Method} & \multicolumn{2}{c}{\textbf{Segmentation Stage}} & \textbf{Mean (CI) [min]} \\ \hline

\multirow{3}{*}{GrandQC}
    & \multicolumn{2}{c}{Tissue vs Background} & 0.01 (0.00, 0.01) \\
    & \multicolumn{2}{c}{Artifact}             & 0.77 (0.75, 0.79) \\
    & \multicolumn{2}{c}{All}                  & 0.78 (0.76, 0.80) \\ \hline

\multirow{7}{*}{HistoKit}
    & \multicolumn{2}{c}{Tissue vs Background} & 4.16 (4.15, 4.16) \\
    & \multirow{5}{*}{Artifact} & Overlap 0\%   & 0.41 (0.39, 0.43) \\
    &                            & Overlap 25\%  & 0.51 (0.49, 0.53) \\
    &                            & Overlap 50\%  & 0.75 (0.71, 0.79) \\
    &                            & Overlap 75\%  & 2.02 (1.85, 2.19) \\
    &                            & Overlap 90\%  & 10.71 (9.36, 12.05) \\
    & All                        & Overlap 75\%  & 6.18 (6.01, 6.35) \\ \hline

HistoQC          & \multicolumn{2}{c}{All} & 0.32 (0.31, 0.33) \\ \hline
HistoROI         & \multicolumn{2}{c}{All} & 0.14 (0.13, 0.15) \\ \hline
Patil et al.     & \multicolumn{2}{c}{All} & 0.25 (0.24, 0.26) \\ \hline
Schreiber et al. & \multicolumn{2}{c}{All} & 0.20 (0.19, 0.20) \\ 
\hline
\end{tabular}%
}

\end{table}

\subsection{Modification and Tuning of HistoKit Algorithm}
\label{sec:modification}

To further improve the segmentation quality of the most accurate artifact detection method, HistoKit, we introduced a simple post-processing step. The proposed method was motivated by several observations about artifact morphology and co-occurrence. Air bubbles were frequently accompanied by out-of-focus regions, likely due to local blurring of the surrounding tissue. Moreover, these two artifact types usually form continuous regions without small holes. In addition, because the GrandQC model (on which HistoKit is based) also predicts the background class during artifact segmentation, small holes may appear within tissue regions. Moreover, small tissue areas may remain adjacent to artifacts when the model misclassifies artifacts. The proposed post-processing procedure was therefore designed to improve segmentation of these two classes, while leaving all other predicted artifacts unchanged.

The area threshold for filling small holes and removing small objects was set to 0.025\% of the total pixels in the WSI mask at 10x magnification. Defining the threshold relative to the image area ensured that the post-processing criteria remained proportional to the spatial resolution and dimensions of the analyzed slide.

First, the mask assigned to the Edge class was processed using a morphological opening operation with a disk-shaped structuring element of radius 5 pixels. This operation removed small and irregularly shaped regions. Subsequently, all holes enclosed within the resulting mask regions were filled using a binary hole-filling operation, thereby producing compact Edge regions. Next, the Out-of-Focus mask was processed similarly. A morphological opening operation with a disk-shaped structuring element of radius 5 pixels was applied, followed by the removal of small holes using the predefined area threshold. The processed Edge and Out-of-Focus masks were then combined, and small holes within their union were filled using the same area threshold. This operation removed small gaps occurring between adjacent Edge and Out-of-Focus regions. Pixels introduced by this operation that did not belong to the Out-of-Focus mask were assigned to the Edge class.

The Tissue mask was processed independently by filling small holes and removing isolated tissue regions according to the predefined area threshold. This reduced small discontinuities and spurious tissue predictions while preserving larger tissue areas. 

Finally, the processed binary masks were recomposed. Pixels belonging to classes not modified by the post-processing procedure were preserved from the original segmentation, and those originally classified as Edge or Out-of-Focus that were removed during morphological processing were reassigned to the Tissue class. The processed segmentation mask was reconstructed by combining the Tissue, Out-of-Focus, and Edge masks in this order. In particular, Edge regions representing air bubbles were placed as the final layer, reflecting the fact that such artifacts typically occur on top of the tissue section and may be surrounded by Out-of-Focus areas.

The project repository provides the complete implementation of the proposed post-processing procedure.

    \label{fig:postprocessing}
    \label{fig:postprocessing}
Additionally, the impact of overlap and results aggregation mode was tested. For the Gaussian mode, four different Gaussian kernel standard deviations were tested: 0.75, 1.25, 1.75, and 2.25. Supplementary Figures S3-S5 present heatmaps showing the median score for each dataset across all parameter combinations. In general, larger overlap led to better results. There are only slight differences between aggregation modes. Because processing time is significantly longer and the differences between 75\% and 90\% overlap are small, we recommend using a slightly smaller patch overlap. Statistical analysis using the paired Wilcoxon test and matched-pairs biserial rank
correlation coefficient effect size showed that the proposed post-processing significantly increased Dice and sensitivity for datasets A and B (Supplementary Figure S6). Differences in FDR were not significant. For Dataset C, no statistically significant differences were noted after post-processing. Figure \ref{fig:postprocessing} shows the selected images before and after post-processing. For dataset A, most improvement came from closing holes in tissue. The largest decrease in Dice for dataset B was due to foreign objects in the form of dental floss classified as air bubble edges. Closing these areas increased misclassified pixels. For dataset C, the proposed approach closed holes in the air bubble, and differences for the worst Dice result are hardly visible.

\begin{figure*}[h!]
    \centering
    \includegraphics[width=1\linewidth, trim=0 0 4cm 0, clip]{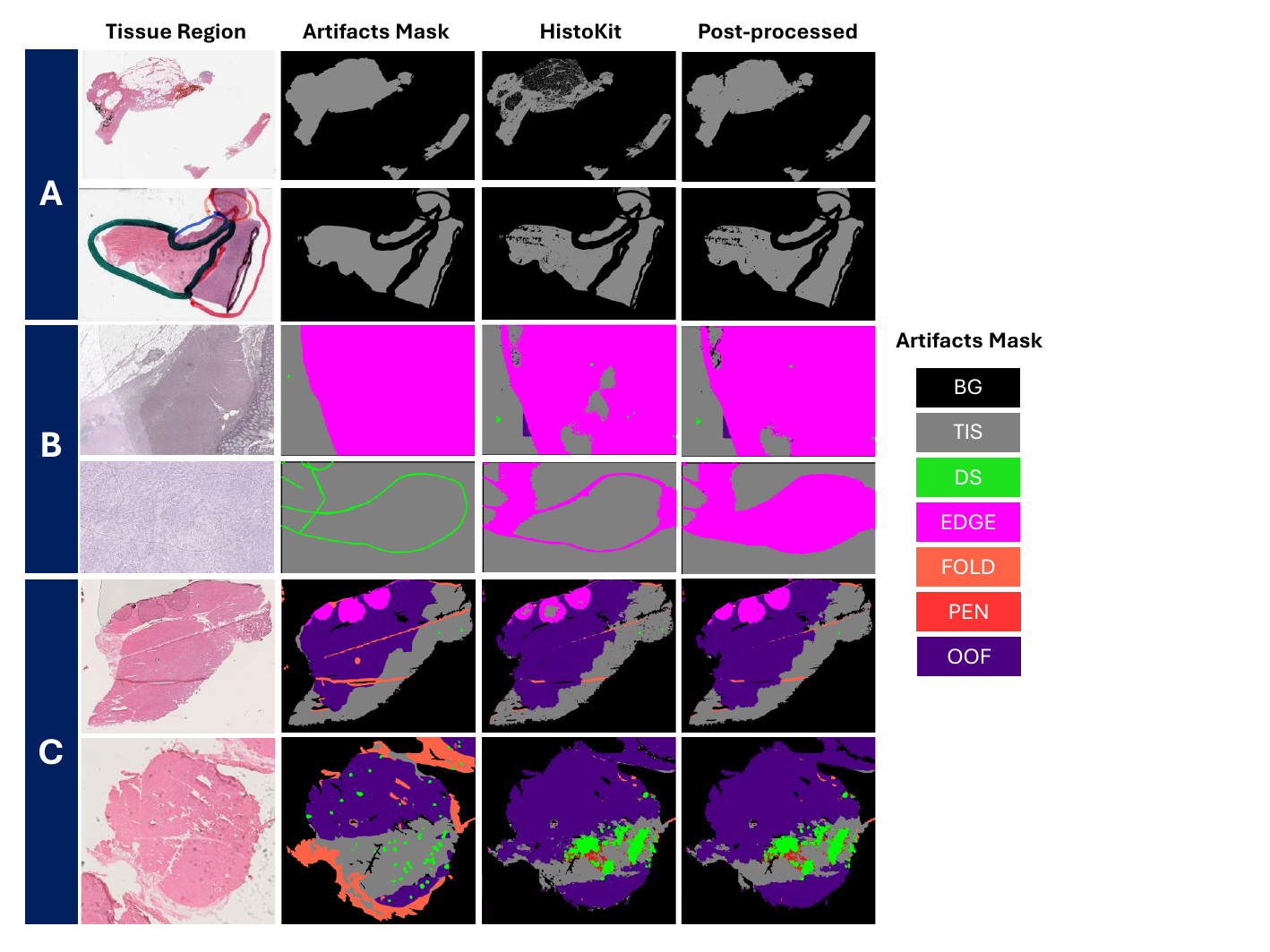}
    \caption{Selected images before and after post-processing HistoKit segmentation masks. From left to right: tissue region, ground-truth artifact segmentation, HistoKit segmentation without post-processing, and HistoKit segmentation after the proposed post-processing.}
    \label{fig:postprocessing}
\end{figure*}

\section{Discussion}

The results showed that HistoKit and GrandQC, both trained on large, diverse datasets, outperformed methods based on thresholding and hand-crafted image features. They yielded superior results for both artifact-free tissue segmentation and segmentation of individual artifact classes. The patch aggregation strategy employed by HistoKit also effectively mitigated distortions at patch boundaries, substantially improving segmentation performance compared with GrandQC. Furthermore, the proposed post-processing approach, which uses the characteristic properties of individual artifact types, improved segmentation quality in most cases. However, HistoKit requires a longer processing time than GrandQC. Therefore, users should consider the trade-off between computational efficiency and segmentation accuracy when selecting the most appropriate method for their application.

It is important to note that the GrandQC artifact segmentation module is included in widely used libraries such as TIA Toolbox v2.1.3 \citep{Pocock2022} and TRIDENT \citep{zhang2025standardizing}, which allow the use of the produced masks to extract patches for downstream tasks, including cell detection, tumor region segmentation, or feature extraction using widely used foundation models. Many of them use magnifications above 10x and typically employ smaller patch sizes, such as $256\times256$ pixels or $224\times224$ pixels. This means a single patch incorrectly predicted by GrandQC, measuring $512\times512$ at 10x magnification, translates into 16 patches of size $ 256\times256$ at 20x magnification and 64 patches at 40x magnification. More precise segmentation produced by HistoKit can significantly affect results from algorithms that directly use the artifact mask to exclude problematic regions.

Our study showed that reporting only the average Dice Index across the entire dataset is insufficient to demonstrate one method's superiority over another, largely because the metric distributions are highly skewed. To identify potential issues, we need detailed statistical comparisons and analysis of both poor and strong cases on several datasets. This underscores the limitations of tools such as SlideGuard \cite{KacGab_SlideGuard_MICCAI2026}, which provide users only average Dice, precision, and recall metrics calculated across the entire dataset. Moreover, the current version of the benchmark does not provide information regarding method performance for specific artifact types. Since annotation masks are not publicly available, it is not possible to identify cases where the tools perform poorly or to verify the quality of the available annotations. Furthermore, the literature review indicates that many artifact detection methods are often not publicly available. Even when project code is accessible, it is often not maintained, and authors often do not provide the model weights. These challenges significantly hinder the ability to conduct benchmarks and to identify the strengths and weaknesses of a given method, thereby reducing confidence in its reliability and real-world usefulness.

The model used by the HistoKit and GrandQC tools can also effectively segment common artifact types in fresh-frozen tissue samples not included in the training dataset, suggesting a potential for generalization across different tissue types and storage methods. However, the five main artifact classes identified in this study do not encompass all possible artifact types that may occur in fresh-frozen tissue sections, such as freezing and thawing artifacts, as none of the evaluated tools was specifically designed to segment them. Thus, preparation of a new annotated dataset or annotation of freezing and thawing artifacts in existing data is still needed.

This study has several limitations. This study did not evaluate how improvements in segmentation quality metrics affect algorithm performance in subsequent tasks, such as cell segmentation or tumor region identification. Future research should investigate how the accuracy of artifact segmentation influences these downstream algorithms. Additionally, the analysis was limited to datasets containing only HE-stained tissues. Assessing algorithm performance on tissue specimens stained with different stains beyond those present in the training datasets would provide insight into the generalization capabilities of the evaluated methods. However, such comparisons remain challenging due to the lack of publicly available datasets with ground-truth annotations.

\section{Conclusions}

This study compared publicly available methods for segmenting artifacts in HE-stained histopathological images across three diverse datasets, evaluating their ability to distinguish artifact-free tissue regions from artifacts and background and to segment different artifact classes. Moreover, a new dataset with precise artifact annotations was released to facilitate benchmarking of future methods. The findings demonstrate that evaluating segmentation algorithms on larger tissue regions, rather than single patches, is crucial for assessing their effectiveness in reducing disturbances at tissue patch edges. Incorporating simple post-processing based on morphological operations that account for the characteristic shape of artifacts further enhanced the performance of the leading algorithm for most datasets. This comprehensive comparison offers valuable insights for future research by evaluating methods in terms of implementation availability, artifact segmentation quality, and processing speed.

\section{Funding}

This work was partially financed by the CoMPaSS-NMD project, “Computational Models for New Patient Stratification Strategies of Neuromuscular Disorders”, funded by the European Union under the Horizon Europe programme, grant agreement number 101080874.

\section{Author Contributions}

\textbf{Julia Merta}:  Writing --- original draft, Visualization, Validation, Software,  Methodology, Investigation, Formal analysis, Data curation. \textbf{Daniel Mendelsohn}: Writing –-- review \& editing, Data curation. \textbf{Benedikt Schoser}: Writing –-- review \& editing, Data curation. \textbf{Joanna Polanska}:  Writing –-- review \& editing, Project administration, Funding acquisition. \textbf{Michał Marczyk}: Writing – review \& editing, Supervision, Project administration, Methodology, Conceptualization.

\section{Declaration of generative AI and AI-assisted technologies in the manuscript preparation process}

During the preparation of this work, the authors used Grammarly to check and correct spelling and grammar in the manuscript. The ideas, methodology, and analyses
presented are the authors’ own. After using this tool, the authors reviewed and edited the content as needed and take full responsibility for the content of the publication.

\section{Author Competing Interests}

The authors declare no competing interests.

\section{Code Availability}

Scripts used to perform statistical analysis and visualizations have been made available in a remote repository on GitHub: \url{https://github.com/julimer228/BenchmarkingArtifactSegmenationHE}

\section{Data Availability Statement}

Masks from Dataset A can be accessed from the official tutorial repository published by \citet{Berman2023SliDL}: \url{https://github.com/markowetzlab/slidl-tutorial/}. WSIs for Dataset A from the CAMELYON16 Dataset can be downloaded from the Grand Challenge platform \url{https://camelyon16.grand-challenge.org/}, while those from the TCGA Database are available at \url{https://portal.gdc.cancer.gov/}.

The complete GrandQC test dataset is publicly available via Zenodo: \url{https://zenodo.org/records/14039591}.

Dataset C and the corresponding manual annotations will be made publicly available via Zenodo upon publication of the manuscript: \url{10.5281/zenodo.23325461}.

\renewcommand\refname{References}
\bibliographystyle{elsarticle-harv} 
\bibliography{localbibliography}

\end{document}